\documentstyle[twocolumn,epsf]{jpsj}
\def \virg{\;\;,}
\def \point{\;\,.}
\def \kf{k_{\rm F}}
\def \vf{v_{\rm F}}

\def \d{{\rm d}}

\def \V2c{V_{\rm 2c}}

\def\ggs{\buildrel\textstyle > \over {\hbox{\raise0.2ex\hbox{$\sim$}}}}
\def\lls{\buildrel\textstyle < \over {\hbox{\raise0.2ex\hbox{$\sim$}}}}
\def\gsim{\,\lower0.75ex\hbox{$\ggs$}\,}
\def\lsim{\,\lower0.75ex\hbox{$\lls$}\,}
\def\runtitle{
Correlation Effect on Peierls Transition 
  }
\def\runauthor
{
 Muneo {\sc Sugiura}   
 and   
 Yoshikazu {\sc Suzumura} 
}

\title{   
Correlation Effect on Peierls Transition 
}

\author{
 Muneo {\sc Sugiura} 
\footnote{E-mail: sugiura@edu2.phys.nagoya-u.ac.jp} 
   and Yoshikazu  {\sc Suzumura} 
\footnote{E-mail: e43428a@nucc.cc.nagoya-u.ac.jp}
}

\inst{
Department of Physics, Nagoya University, Nagoya 464-8602 \\ 
}

\recdate{\hspace{35mm}}

\abst
{
The effect of correlation on   Peierls transition, which is accompanied 
 by  a dimerization, $t_d$, of  a bond alternation for  
  transfer energy,  has been examined 
 for a half-filled one-dimensional 
 electron system with  on-site repulsive interaction ($U$). 
 By applying  the renormalization group method to the interaction of 
  the bosonized  Hamiltonian,  the dimerization has been calculated 
 variationally and self-consistently 
 with a fixed  electron-phonon   coupling constant ($\lambda$)
  and it is shown  that  $t_d$   takes a maximum as a function of  $U$.
 The  result is examined in terms of  charge gap and spin gap 
  and is compared with that of the  numerical 
  simulation  by Hirsch 
  [Phys. Rev. Lett {\bf 51} (1983) 296]. 
 Relevance to the spin Peierls transition in 
  organic conductors is discussed.  
}

\kword
{
 Peierls transition, Hubbard model, dimerization, 
 renormalization, charge gap, spin gap 
  }

\begin{document}
\sloppy
\maketitle

Peierls transition  for  a one-dimensional 
   half-filled electron system coupled with  phonon
   has been studied extensively  since  
  the   Su-Schrieffer-Heeger 
 model was proposed for   the quasi-one-dimensional 
     conductor polyacetylene with a bond alternation.
\cite{SSH} 
 The  role of correlation in such a model has been  examined 
  by introducing  an on-site repulsive interaction, $U$. 
 The calculation using the  Hartree-Fock approximation leads to    
 the  Peierls state, which  exists  only for $U$ smaller than 
  a critical value of the order of the band width.
\cite{Subbaswamy,Kivelson}
This  study  has been further developed   by 
  taking into account   a one-dimensional quantum fluctuation. 
 The numerical simulation exhibits an enhancement of dimerization 
 in the presence of   repulsive interaction.
\cite{Mazumdar,Hirsch,Dixit,Soos,Baeriswyl,Hayden} 
A notable finding
 is  that  the dimerization takes a maximum 
 at a value of $U$ being nearly  the band width
\cite{Hirsch,Baeriswyl,Hayden}
 and that the  charge gap becomes much larger than the  dimerization gap.
\cite{Hirsch} 
Regarding  the case of weak coupling, 
 the   analytical  method  of  a  renormalization group (RG) has also  
  exhibited  the enhancement of $t_d$ as a function of $U$.
\cite{Horovitz}
The effect of  
      finite phonon frequency
 has been  explored 
 using  the  RG method  based on the bosonization.  
\cite{Voit_Schulz,Voit,Yonemitsu}
  It has been shown that 
  the  half-filled  case leads to    a competition 
    between the state with 
   both spin and charge gaps and the state with only a charge gap 
     on the plane of  the phonon frequency and the electron-phonon 
   coupling  constant.
\cite{Yonemitsu}  
In contrast, the case of strong coupling with large $U$ 
 has been examined 
   in terms of   the opposite  approach, i.e.,    
       the expansion of $1/U$) and  
   mapping the electron system  into a spin 1/2 chain system,
   which  leads to the  spin Peierls transition. 
 The successful treatment of the quantum fluctuation leads to 
 the occurrence of 
   the spin Peierls transition  for an arbitrary magnitude of 
 the electron-phonon coupling constant. 
\cite{Cross,Nakano,Inagaki}  
However, it is not clear  why the optimum condition for 
 the Peierls transition is given by 
  the intermediate coupling  of  $U$.

In  the present paper, the unconventional role  of $U$ on the Peierls
  state is  examined by applying the RG 
   method to  the bosonized Hamiltonian. 
 It is demonstrated  that 
  a maximum of the dimerization occurs for  $U \simeq 2t$ 
 ($4t$ is the band width)
  for weak  electron-phonon coupling ($\lambda$),
  and that
  the charge gap is well separated from  the spin gap 
    at the maximum. 
 Furthermore,   
 we discuss the relevance of this  to 
   the spin Peierls transition in an  organic conductor,
 (TMTTF)$_2$PF$_6$ salt.
\cite{Yamaji,Bechgaard}

We consider a one-dimensional half-filled Hamiltonian given by
\cite{SSH,Subbaswamy,Kivelson} 
\begin{eqnarray}
H  = & &  \,   \sum_{j} \Bigl[ - \,
  \sum_{\sigma} \,
        (  t  -  (-1)^j  t_d ) 
        (  c^\dagger_{j,\sigma} c_{j+1,\sigma}  +  h.c.)
              \nonumber \\
   & &   
 +  U  \, n_{j,\uparrow}  n_{j,\downarrow} 
      +   C  t_d^2/2  \Bigr] \virg  
\label{eq:Hamiltonian}
\end{eqnarray}
where  $c^\dagger_{j,\sigma}$ denotes a creation operator of a conduction 
 electron with spin $\sigma (= \uparrow, \downarrow)$ 
 at the lattice site $j$. 
 The first term is the  kinetic  energy, 
 where $t$ is the uniform transfer energy.
  The quantity,  $t_d$, which   denotes the dimerization 
    due to  Peierls distortion, 
 is determined so as to minimize the total energy. 
The $U$ term  with 
  $n_{j, \sigma} = c_{j,\sigma}^{\dagger} c_{j,\sigma}$
 denotes an on-site repulsive interaction, 
 and the last term with a constant $C$  is  the elastic energy 
   for the distortion. 
  
Applying the bosonization method,
\cite{Luther_Peschel,Mattis} 
 eq.(\ref{eq:Hamiltonian}) for the half-filled  case is 
 rewritten as  
\begin{eqnarray}
 H  &=& 
     + \; \frac{v_\rho}{4\pi} \int dx\;
  \left[\:\frac{1}{K_\rho} 
    (\partial_x \theta_+)^2+K_\rho (\partial_x \theta_-)^2\: \right] 
         \nonumber \\
            & & \nonumber \\ 
  & & 
     + \;\frac{v_\sigma}{4\pi} \int dx\;
  \left[\:\frac{1}{K_\sigma} 
       (\partial_x \phi_+)^2+K_\sigma (\partial_x \phi_-)^2\: \right]
        \nonumber \\
            & & \nonumber \\
  & & + \; \frac{v_F}{2\pi\alpha^2} \, 
         \int dx \, 
     \Bigl[  y_{1/2} \, \cos2\theta_+ 
           +   y_\sigma \, \cos2\phi_+   
 \nonumber \\
   & &    
      -   y_W \, \sin\theta_ + \cos\phi_+ 
           +  (1/8 \lambda) y_W^2  \Bigr] 
               \virg 
\label{eq:PhaseH} 
\end{eqnarray}
 where $\theta_{\pm}(x)$ and  $\phi_{\pm}(x)$ denote 
 phase variables for  the charge and spin fluctuation, respectively,
\cite{Suzumura} 
 and
$
 [ \theta_+ (x)  ,  \theta_- (x^\prime) \, ] =  
   [ \phi_+ (x)  ,  \phi_- (x^\prime) \, ] =  
    {\rm i} \, \pi \, {\rm sgn}(x - x^\prime) \label{eq:k} \point 
$
The $y_{1/2}$ term and the $y_W$ term represent  the Umklapp scattering 
 and the dimerization,  respectively. 
The coefficients in  eq.(\ref{eq:PhaseH}) 
   are given by 
$ 
  K_\rho   
      =  (1 + \tilde{U})^{-1/2}  ,
$
$
  K_\sigma  
       = (1 - \tilde{U})^{-1/2} ,  
$
$
 v_\rho 
   = \vf  (1 \, + \, \tilde{U})^{1/2} , 
$
$
  v_\sigma 
      =  \vf  (1 \, - \, \tilde{U})^{1/2} ,
$
$
 y_{1/2} = y_{\sigma} = \tilde{U}, 
$ 
$
 y_W =  8 \alpha t_d /\vf  , 
$
$
1/\lambda = \pi \vf C/8 a, 
$
$ 
\vf = 2 t a  \sin \kf a  (= 2 t a), 
$
$\kf=\pi/(2a)$
and  $\tilde{U} = U a / (\pi \vf)$.
 The quantity $\alpha$ is a cutoff parameter of the order of the 
 lattice constant, $a$;  the  ratio of $\alpha$ to $a$ 
  will be  evaluated later. 
The quantity  $y_W$  is determined 
  by minimizing the total energy, and  the resultant 
  self-consistency equation is written as  
\begin{eqnarray}
\frac{1}{4 \lambda} \: y_W \, 
 =  \langle \, \sin\theta_+ \, \cos\phi_+ \, \rangle 
 \left( \equiv  \, \Delta \label{eq:sc} \right) \virg 
\label{eq:SCE}
\end{eqnarray}
where $y_W = 4 t_d \alpha/(t a)$.
The quantity  $\Delta$, which 
  denotes an order parameter for the dimerization, 
  is calculated self-consistently in terms of eq.(\ref{eq:PhaseH}). 
Equation (\ref{eq:PhaseH}) is examined using  
 the RG  method with 
 a scaling $\alpha \rightarrow \alpha (1+ d l)$.    
 Within the lowest order of perturbation, 
   RG  equations for the  coupling constants  are   derived as
\cite{Solyom_adv,Giamarchi,Voit_Schulz,Voit,Yonemitsu,Tsuchiizu} 
\begin{eqnarray}
 \frac{d}{dl}\:K_\rho(l) \, 
  &=& -\,\frac{1}{2}\;y_{1/2} ^2(l) \; 
    K_\rho^2(l)
           - \,\frac{1}{16}\;y_W ^2(l) \; K_\rho^2(l) \, , 
      \nonumber \\
 \frac{d}{dl}\:G_\sigma(l) \, 
  &=& \, - \, y_\sigma^2(l) -\;\frac{1}{8} \, y_W ^2(l) \, , 
     \nonumber \\
 \frac{d}{dl}\:y_{1/2}(l) \, 
  &=& \, [ \, 2 \, - \, 2 \, K_\rho(l) \, ]\,
      y_{1/2}(l)\;     + \;\frac{1}{8} \, y_W^2(l) \, , 
      \nonumber \\
 \frac{d}{dl}\:y_\sigma(l) \,
   &=& \, - \, G_\sigma(l) \, y_\sigma(l) 
        - \;\frac{1}{8} \, y_W ^2(l) \, , 
       \nonumber \\
 \frac{d}{dl}\:y_W(l) \, 
   &=& \, \left[ \, \frac{3}{2} \, 
       - \, \frac{1}{2} \, K_\rho(l)\, 
        - \, \frac{1}{4} \, G_\sigma(l) \, \right]\,y_W(l) 
                       \nonumber \\
  & & \hspace{0cm}     + \; \frac{1}{2}\;y_{1/2}(l) \, y_W(l)
          - \; \frac{1}{2}\;y_\sigma(l) \, y_W(l) \, ,
          \label{eq:RGE}
\end{eqnarray}
where 
 $ G_{\sigma}(l) = 2 (K_{\sigma}(l) - 1) $ 
 and initial values at $l =0$  are given by 
  coefficients of eq.(\ref{eq:PhaseH}). 
The relevance of  $y_{1/2}$, $y_{\sigma}$ and $y_{W}$
 corresponds to the appearance 
  of charge gap, spin gap  and dimerization,  
 respectively. 

For calculating the r.h.s. of eq.(\ref{eq:SCE}), we use the fact that 
 the derivative of $ \langle 
 \sin \theta_+ \cos \phi_+ \rangle $ with respect to $y_W$ 
  can be expressed  in terms of the  response function 
 given by
 \begin{eqnarray}
  R(|\vec{r}_1 - \vec{r}_2|) = 
   \left< T_\tau O_d(\vec{r}_1)  O_d(\vec{r}_2) \right> \virg 
\end{eqnarray}
 where 
 $
O_d(\vec{r})
 (
 \equiv   \sin\theta_+(\vec{r}) \cos\phi_+(\vec{r})
  = - (\pi  \alpha/4a)
 \sum_{\sigma}
     (-1)^{j} c^\dagger_{j,\sigma}(\tau) 
                     c_{j+1,\sigma}(\tau) + h.c. )
$
 denotes an operator for  dimerization.
 The vector $\vec{r}$ consists of   space $x(=ja)$ and  imaginary 
time $\tau$   
 while  $T_{\tau}$ is the time-ordering operator. 
Using the  RG method,  $R(t_d,l) (= R(|\vec{r}|))$  is calculated as
\cite{Giamarchi_Schulz} 
\begin{eqnarray} 
R(t_d, l) 
 & = &   \frac{1}{4} \, \exp \, \Bigl[ \, - \int_{0}^{l} \d l^\prime 
    \, ( K_\rho(l^\prime) 
               \nonumber \\
              & & 
       + K_\sigma(l^\prime) - y_{1/2}(l^\prime) 
        + y_\sigma(l^\prime) ) \, \Bigr] \virg 
                        \hspace{0.3 cm}  (l \le l_c)                                                          \nonumber   \\
R(t_d, l) 
  & = & 
   \frac{1}{4} \, \exp \, \Bigl[ \, - \int_{0}^{l_c} \d l^\prime 
     \, ( K_\rho(l^\prime) 
               \nonumber \\
              & &  
   + K_\sigma(l^\prime) 
    - y_{1/2}(l^\prime) + y_\sigma(l^\prime) ) \, \Bigr] 
                                                 \nonumber \\  
  & &  \times  \exp \, \Bigl[ \, - \int_{l_c}^{l} \d l^\prime  
           \, ( K_\sigma (l^\prime)+ y_\sigma(l^\prime) ) \, \Bigr] 
                                 \virg        \hspace{0.3 cm}      (l>l_c) 
                                \nonumber \\ 
                        \label{eq:Response}  
\end{eqnarray}
where 
 $l = \ln ( [x^2 + (v_F \tau)^2]^{1/2}/\alpha)$. 
 In eq.(\ref{eq:Response}),  
 we introduced   $l = l_c$ with  $y_{1/2}(l_c) = 1$, from which 
  the charge gap is calculated.  
 The response function for  $l > l_c $ is calculated by estimating   
 $ \langle \sin \theta_+ \rangle $  in the presence of  the charge gap.  
  Although one expects  
  $ R(\infty) \rightarrow \Delta^2 $, 
  we use the  following method to estimate $\Delta$ 
    within  the  present  RG equation,
      which leads to $R(l)$ taking a minimum  at $l=l_m$. 
 Since   $R(l)$ as a function of $l$  becomes 
  invalid for $l > l_m$, 
   we use a self-consistency  equation 
     for $\Delta (= \Delta (t_d) )$ given by 
\begin{eqnarray}
\Delta(t_d) &=& \int_{0}^{t_d} \, 
                             \d t^\prime_d \,
   \int_{0}^{l_m(t_d^{\prime})}  \d l^\prime \, e^{2 l^\prime}
\left\{ \, R(t^\prime_d,l^\prime) \,- \, 
  \Delta^2(t^\prime_d) \, \right\} \point
          \label{eq:Delta} 
\end{eqnarray}

We assume $t=1$ in the following numerical calculation, 
 if there is no confusion.  
 For estimating the r.h.s. of eq.(\ref{eq:Delta}), we  
   calculate  RG  eq.(\ref{eq:RGE}). 
 When $l$ increases with fixed $U (>0)$, one finds 
   the decrease of  $K_{\rho}(l)$, $G_{\sigma}(l) $ and $y_{\sigma}(l)$ 
     and the increase of $y_{1/2}$ and $y_W$.  
Behaviors at large  $l $ are classified into two groups  
  depending on the magnitude  of $y_W(0)$.
For small  $y_W(0)$,   $y_{1/2}(l)$ is   larger   than $y_W(l)$,  while   
 $y_{1/2}(l)$ becomes  smaller than $y_W(l)$ with increasing  $y_W(0)$. 
 The former case denotes the Peierls state in the presence of 
 a  well-developed charge gap since  
   the relevance of $y_{1/2}$  ($y_W$) corresponds to the 
   formation of the charge gap (the formation of 
 both charge and spin gaps    due to  the Peierls  distortion).
\begin{figure}[tb]
\begin{center}
\vspace{2mm}
\leavevmode
\epsfysize=7.5cm\epsfbox{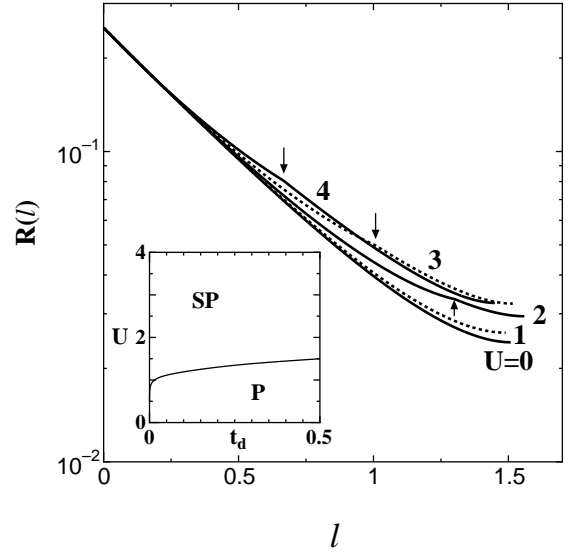}
 \vspace{-3mm}
\caption[]{
$l$-dependence of $R(l)$ of eq.(\ref{eq:Response}) 
  for   
 $t_d = 0.1$, and  $U$ =0, 1, 2, 3 and 4 
 where   $R(l)$ decreases and  takes a minimum at $l=l_m$. 
 $ t = $ 1 and $\alpha/a  =$ 1.52.   
The arrow denotes $l=l_c$ given by   $y_{1/2}(l_c)=1$. 
In the inset, the spin Peierls (SP) region and  
 the Peierls (P) region  correspond  to 
  $ l_c<l_m$ and  $l_c > l_m$, respectively, where 
 the boundary denotes a crossover  between these two regions.  
 }
\end{center}
\end{figure} 

 The response function is calculated by substituting 
  the solution of RG equations into eq.(\ref{eq:Response}).
 In Fig.1, the $l$-dependence of $R(l)$ is shown for $t_d =0.1$ 
    and certain values  of $U$. 
  The results are shown in  the  region with $l < l_m$,
     where $R(l_m)$ is  a minimum  of $R({l})$.
  The long range order of the Peierls state  is related to  
     the relevance of $y_W$, which leads to   
       the appearance of   $l_m$  in 
         the present approximation of the RG  equations.   
 The arrow for $U =$ 2, 3 and 4 denotes the location 
     at   $l = l_c$ where $y_{1/2}(l_c) = 1$.   
 From a comparison of $l_c$ with  $l_m$, 
      one finds  a distinction between  case I (  $l_m > l_c$ )
            and  case II ( $l_m < l_c$).  
 The charge gap is already developed prior to  obtaining 
    the  Peierls state in  case I.  
 The inset depicts   
   the spin Peierls region (case I) and the Peierls region (case II) 
     on the plane of   $U$ and $t_d$. 
  The charge gap dominates for $U \gsim 1.5$ and $t_d < 0.5$. 
 The solid curve shows the boundary as 
 a crossover  between these two regions,  where 
  the state moves  continuously.  
 

 An estimation of  the  ratio of  $\alpha / a $ is needed for  
 the  calculation of   $\Delta$ of  eq.(\ref{eq:Delta}) 
  as a function of $t_d$ and $U$.    
 When  $U$ = 0, 
   $\Delta$ becomes  equal to the order parameter 
    of the conventional Peierls distortion and thus 
 the r.h.s. of eq.(\ref{eq:SCE}) 
       can be calculated rigorously as
 $ \Delta \rightarrow 
 \Delta_{\rm Peierls} =  
  (t_d \alpha / a)                
    \int_{0}^{\pi} \d z \, 
                   (\sin z)^2/[(2  \cos z)^2 \, 
                    + \, (2 t_d \sin z)^2]^{1/2} \virg 
$
 which leads to 
 $  (t_d \alpha / a) \times  {\rm ln} (1.4/t_d)$ for small $t_d$.  
By comparing   $\Delta_{\rm Peierls}$ with 
that  obtained from eq.(\ref{eq:Delta}),  we obtain 
  $\alpha / a \simeq 1.5 \pm 0.02$ for small $t_d ( \lsim 0.2)$.
     
\begin{figure}[htb]
\begin{center}
 \vspace{2mm}
 \leavevmode
\epsfysize=7.5cm\epsfbox{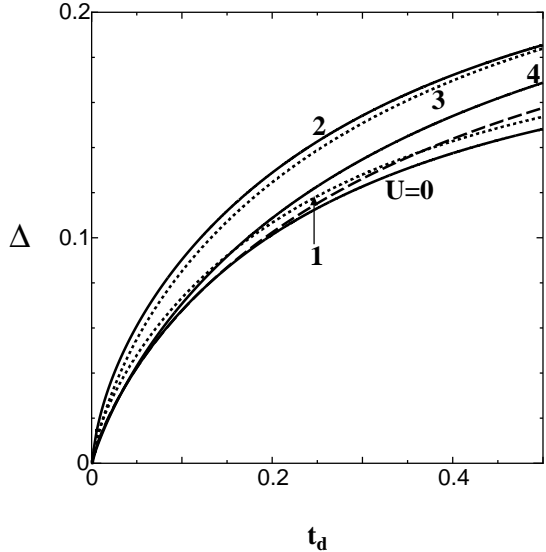}
 \vspace{-3mm}
\caption[]{
$t_d$-dependence of $\Delta$  calculated self-consistently by using 
  eq.(\ref{eq:Delta})   for $U =$ 0, 1, 2, 3 and 4. 
 The dashed curve denotes $\Delta_{\rm Peierls}$.
 }
\end{center}
\end{figure} 
In terms of the response function, 
  $\Delta$ is calculated self-consistently  from eq. (\ref{eq:Delta}).
In Fig. 2, $\Delta$  is shown  as a function of $t_d$  
    for certain values  of $U$.
The dashed curve denotes 
 $\Delta_{\rm Peierls} $ which coincides well with 
 that of $U=0$  (solid curve) for $t_d \lsim 0.2$.
 With increasing $t_d$,  $\Delta$ for  fixed $U$ increases monotonically.
 The self-consistent result of $\Delta$ obtained from 
 eq. (\ref{eq:Delta})  is smaller than  $R(l_m)^{1/2}$, 
 $e.g.$, 
 $\Delta / R(l_m)^{1/2} =$ 0.43 ,0.45  ,0.52, 0.47   and  0.39  for 
 $U =$ 0 ,1, 2, 3 and 4.
This is reasonable since 
   the lowest order RG equation with the relevant coupling 
  overestimates  the response function for large $l$.  
 From the comparison of $\Delta$ for $U = $ 2 with  that for $U =$ 3, 
 it is found that $\Delta$ as a function of $U$ (\gsim 2) 
 decreases, indicating the existence of an optimum value of $U$ 
  for the Peierls state. 

\begin{figure}[tb]
\begin{center}
 \vspace{2mm}
 \leavevmode
 \epsfysize=7.5cm\epsfbox{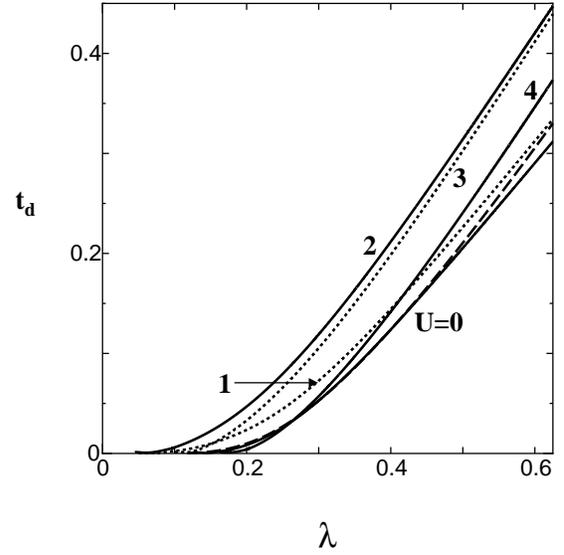}
 \vspace{-3mm}
\caption[]{
$\lambda$-dependence of $t_d$, which is calculated from 
  eq.(\ref{eq:SCE}),  for $U =$ 0, 1, 2, 3 and 4 where 
 the dashed curve denotes $\Delta_{\rm Peierls}$ 
 corresponding to the conventional calculation in the absence of $U$.
}
\end{center}
\end{figure}
Based on Fig. 2, we obtain the dimerization, $t_d$, representing  
 the Peierls state, which is calculated 
 using  eq.(\ref{eq:SCE}). 
In Fig. 3, $t_d$
  is shown  as a function of $\lambda$ with certain values  of $U$.
 The dashed curve denotes 
 $t_d (\simeq 1.4 \exp[-1/\lambda])$ obtained  
   from $\Delta_{\rm Peierls}$, 
 which coincides with  $t_d$ in the case of  $U = 0$ (solid curve) 
  for  $0.25 \lsim  \lambda \lsim  0.5$   within the visible scale. 
For small $\lambda$, $t_d$ decreases exponentially 
 but is enhanced sufficiently for $U$ = 1, 2 and 3. 
 We note that 
  $t_d$  takes a maximum around $U \simeq 2$. 
 Such a  $U$-dependence of $t_d$, indicating  a crossover from 
 the regime of weak coupling to that of strong coupling, 
  is elucidated  by examining the relevant term  of  
   $y_W \sin \theta_+ \cos \phi_+$ in  
  eq.(\ref{eq:PhaseH}) and the corresponding   RG equation  
 (the last equation of  eq.(\ref{eq:RGE})). 
 For small $U$,    the $y_W$-term  is enhanced by 
   the Umklapp scattering of the $y_{1/2}$-term, which  leads  to  
      the suppression  of  charge fluctuation ($\theta_+$).  
    For large $U$, both $y_{\sigma}$  and $K_{\sigma }$
      increase the quantum fluctuation for the spin part ($\phi_+$)
        resulting in the suppression of $ <\cos \phi_+ > $
          for the Peierls state. 

\begin{figure}[tb]
\begin{center}
 \vspace{2mm}
 \leavevmode
 \epsfysize=7.5cm\epsfbox{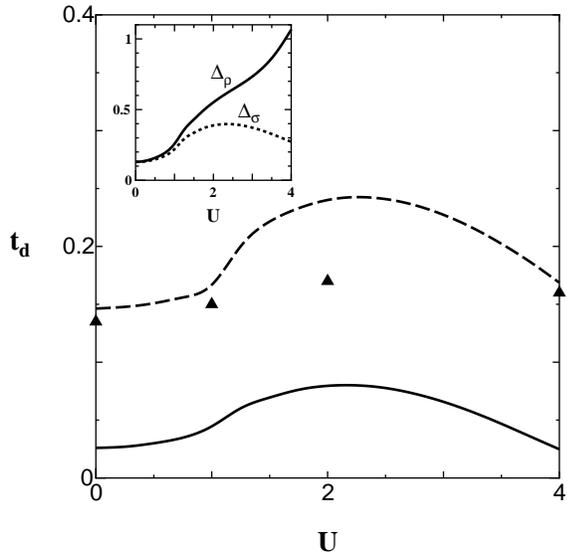}
 \vspace{-3mm}
\caption[]{
Quantity $t_d$ as a function of $U$ for $\lambda =$ 0.25 (solid curve).
  The symbol (triangle) is the result calculated by Hirsch
\cite{Hirsch}, which is compared with the present calculation with 
  $\lambda =$ 0.42 .
The inset shows the charge gap ($\Delta_{\rho}$) and the spin gap 
 ($\Delta_{\sigma}$) for $\lambda = 0.25$. 
}
\end{center}
\end{figure}
 The $U$-dependence of $t_d (=\Delta \lambda a/ \alpha)$ is obtained 
   by examining $t_d$ with fixed $\lambda$ in Fig. 3.  
  In Fig. 4, $t_d$  with $\lambda = 0.25$ 
   is depicted  as a function of $U$ by the solid curve. 
As expected from Fig. 2 and Fig.3, $t_d$ 
 takes a maximum around $U = 2$. 
The  triangle, which represents  the numerical result obtained  by Hirsch,
\cite{Hirsch} 
  is compared with the dashed  curve of 
  the present calculation  with   $\lambda = 0.42$ 
 (dashed  curve)
   corresponding to  the parameter of the numerical simulation.  
 The fact that the dashed curve is larger than the numerical  
  result (triangle) 
   is reasonable  since the present calculation treats the 
   dimerization  ( the bond alternation)   classically 
    while the numerical simulation has been performed 
     in the presence of the quantum fluctuation. 
 We note that there is not much variation of 
   the location of $U$ for the maximum, $e.g.$, 
the maximum appears at $U = 2.2 \pm 0.1$  
 in the interval range of 
    $0.2 \lsim \lambda \lsim  0.5$. 
In the inset, 
 the $U$-dependences of the charge gap ($\Delta_{\rho}$) and 
  spin gap ($\Delta_{\sigma}$)  for $\lambda = 0.25$ are shown 
   by a solid curve and   dotted curve, respectively. 
 These gaps   are calculated by  
  $\Delta_{\rho}/\omega_c  = \exp [- l_c]$ 
  and $\Delta_{\sigma}/\omega_c = \exp [- l_{\sigma}]$ with  
  $| y_{\sigma}(l_{\sigma})| = 1 $,  where  
    $\omega_c ( \simeq 2.37$) is chosen so as to obtain 
    $\Delta_{\rho}$  with  $ t_d = 0$  nearly  equal    
  to the well-known exact  one.
 With increasing $U$, $\Delta_{\rho}$ increases monotonically 
  while $\Delta_{\sigma}$ takes a maximum.
 The maximum of $t_d$ is found to be accompanied   
  by  the separation  of charge gap and spin gap.   
 Thus, it is expected that with  increasing $U$, there is a crossover 
 from the conventional  Peierls state with 
  $\Delta_{\rho} \simeq \Delta_{\sigma}$ to  
    the spin Peierls state with  
 the charge gap being  much larger than the spin gap. 
  
 When  $U$ is  extremely large, 
 the effect of $t_d$ can be examined 
   using the bosonized  phase Hamiltonian
 of   spin 1/2 chain
\cite{Inagaki}
 with the antiferromagnetic exchange energy, $4(t\pm t_d)^2/U$ .
 The calculation of $t_d$, similar to eq.(\ref{eq:SCE}), 
   with only a spin degree of freedom
  shows that   $t_d$ decreases  monotonically 
   with increasing $U$. 
 From comparison of the present result with this limiting one    
   (not shown explicitly),    
   the present calculation   for $U \lsim 4$ in Fig. 3 seems to be 
 reasonably   extrapolated to that of the spin 1/2 chain  
   except for small $t_d$ (and then  small $\lambda$), which requires 
       much  accuracy of numerical evaluation.

Finally,  we discuss  the spin Peierls transition 
 in an  organic conductor,
 (TMTTF)$_2$PF$_6$, which has been observed by magnetic and X-ray 
 experiments.
\cite{Pouget,Dumm} 
This conductor, which indicates the charge gap being much larger than 
 the spin gap
\cite{Bechgaard}
  is often analyzed in terms of a model with spin 1/2 chain 
 although  $U$ is of the order of the band width.
\cite{Mila}
 The conductor has 
  a quarter-filled band with  a dimerization, 
     which may be considered to be   half-filling.  
 However, such an effectively half-filled band strongly reduces    
 the magnitude of  Umklapp scattering, resulting in  
 the suppression of the Peierls state. 
 In fact, 
 the Umklapp scattering is estimated to be   
   $ y_{1/2}= 2 (x_d /(1+x_d^2)) \tilde{U}$, 
 where  $\vf =\sqrt{2}ta$ and  
$x_d$ corresponds to  a dimerization
  for a quarter-filling. 
\cite{Tsuchiizu_Y_S} 
 For  $x_d \simeq  0.1$   corresponding to   TMTTF salt,
\cite{Yamaji}
 we find that $t_d$ as a function of $U$ decreases 
   for  small $U$  while  separation 
 between $\Delta_{\rho}$ and $\Delta_{\sigma}$ appears  for 
 $ U  \gsim 3$.      
Moreover, with decreasing $x_d$ and  fixed $\lambda ( \lsim 0.5)$,  
  $t_d$ decreases strongly  for $U = 4 \sim 6$.  
 When  $U = 5.6$ and   $\lambda = 0.25 (0.5)$, 
 we obtain the dimerization as $t_d =$ 0.036 (0.23), 0.019 (0.17) 
  and $\simeq$ 0 (0.05); 
  the  charge gap as $\Delta_{\rho} =$ 0.31 (1.05), 0.12 (0.73) 
   and $\simeq$ 0 (0.17);   
  the spin gap as  $\Delta_{\sigma} =$ 0.18 (0.77), 0.08 (0.57) 
    and  $\simeq$ 0 (0.15), 
  for  $x_d$ = 0.2, 0.1 and 0, respectively.
Based on such a  consideration, 
 the   spin Peierls state in the organic conductor  
 could be realized   when  Umklapp scattering induced by 
 the electronic correlation 
 becomes large. 

In summary, 
 we have examined the effect of on-site repulsive interaction, $U$,  on 
 the Peierls state with dimerization $t_d$. 
 For small (large) $U$, $t_d$ increases (decreases) 
   where  the charge gap is almost equal to (much larger than) 
   the spin gap. 
 The maximum of $t_d$ indicates a crossover from 
  the weak coupling  regime, in which the Umklapp scattering  
    suppresses  the  charge fluctuation,     
    into a strong coupling regime, in which  
      the quantum spin fluctuation reduces the effect of dimerization.




\begin{thebibliography}{99} 
\def\jo #1#2#3#4{#1 {\bf #2} (#3) #4}  
\def\PS{Physica.\ Scripta.}
\def\AC{Acta.\ Cryst.}
\def\JPSJ{J.\ Phys.\ Soc.\ Jpn.}
\def\JMP{J.\ Math.\ Phys.}
\def\PRB{Phys.\ Rev.\ B}
\def\PRL{Phys.\ Rev.\ Lett}
\def\PTP{Prog.\ Theor.\ Phys.}
\def\JLP{J.\ Low Temp.\ Phys.}
\def\ADV{Adv.\ Phys.}
\def\JPF{J.\ Phys.\ France}
\def\ZP{Z.\ Physik}
\def\SYM{Synth.\ Met}



\bibitem{SSH}
 W.P. Su, J.R. Schrieffer and A.J. Heeger: 
\jo{\PRL}{42}{1979}{1698}.
\bibitem{Subbaswamy}
 K.R. Subbaswamy and M. Grabowski:
\jo{\PRB}{24}{1981}{2168}.
\bibitem{Kivelson}
 S. Kivelson and D.E. Heim:
\jo{\PRB}{26}{1982}{4278}.
\bibitem{Mazumdar}
 S. Mazumdar and S.N. Dixit:
\jo{\PRL}{51}{1983}{292}.
\bibitem{Hirsch}
 J.E. Hirsch:
\jo{\PRL}{51}{1983}{296}.
\bibitem{Dixit}
 S.N. Dixit and S. Mazumdar:
\jo{\PRB}{29}{1984}{1824}.
\bibitem{Soos}
 Z.G. Soos and S. Ramasesha:
\jo{\PRB}{29}{1984}{5410}.
\bibitem{Baeriswyl}
 D. Baeriswyl and K. Maki:
\jo{\PRB}{31}{1985}{6633}.
\bibitem{Hayden}
 G.W. Hyden and E.J. Mele:
\jo{\PRB}{32}{1985}{6527}.
\bibitem{Horovitz}
 B. Horovitz and J. S\'olyom:
\jo{\PRB}{32}{1985}{2681}.
\bibitem{Voit_Schulz}
 J. Voit and H.J. Schulz:
\jo{\PRB}{37}{1988}{10068}.

\bibitem{Voit}
 J. Voit:
\jo{\PRL}{64}{1990}{323}.

\bibitem{Yonemitsu}
 K. Yonemitsu and M. Imada:
\jo{\PRB}{54}{1996}{2410}.

\bibitem{Cross}
 M.C. Cross and D.S. Fisher:
\jo{\PRB}{19}{1979}{402}.
\bibitem{Nakano}
 T. Nakano and H. Fukuyama:
\jo{\JPSJ}{49}{1980}{1679}.

\bibitem{Inagaki}
 S. Inagaki and H. Fukuyama:
\jo{Kotaibutsuri}{20}{1985}{369} [in Japanese]. 

\bibitem{Yamaji}
 T. Ishiguro and K. Yamaji: 
 {\it Organic Superconductors} (Springer-Verlag, Berlin, 1990). 
\bibitem{Bechgaard}
 K. Bechgaard and D. J\'erome: 
\jo{\PS}{T39}{1991}{37}.



\bibitem{Luther_Peschel}
 A. Luther and I. Peschel:
\jo{\PRB}{9}{1974}{2911}.

\bibitem{Mattis}
 D.C. Mattis and E.H. Lieb:
\jo{\JMP}{6}{1965}{304}.

\bibitem{Suzumura}
 Y. Suzumura:
\jo{\PTP}{61}{1979}{1}.


\bibitem{Solyom_adv}
 J. S\'olyom:
\jo{\ADV}{28}{1979}{201}.

\bibitem{Giamarchi}
 T. Giamarchi and H.J. Schulz:
\jo{\JPF}{49}{1988}{819}.

\bibitem{Tsuchiizu}
 M. Tsuchiizu and Y. Suzumura:
\jo{\JPSJ}{68}{1999}{3966}.

\bibitem{Giamarchi_Schulz}
 T. Giamarchi and H.J. Schulz:
\jo{\PRB}{39}{1989}{4620}.


\bibitem{Pouget}
 J.P. Pouget, R. Moret, R. Comes,
 K. Bechgaard, J.M. Fabre and L. Giral:
Mol. Cryst. Liq. Cryst. {\bf 79} (1982) 129.
 
\bibitem{Dumm}
 M. Dumm, M. Dressel, A. Loidl, B.W. Frawel, 
K.P. Starkey  and L.K. Montgomery: 
\jo{\SYM}{103}{1999}{2068}.
\bibitem{Mila}
F. Mila:
\jo{\PRB}{52}{1995}{4788}.

\bibitem{Tsuchiizu_Y_S}
 M. Tsuchiizu, H. Yoshioka and Y. Suzumura:
\jo{\JPSJ}{68}{1999}{1809}.

\end{thebibliography}
\end{document}